\documentclass[final,authoryear,5p,times,twocolumn]{elsarticle}

\usepackage{graphicx}

\usepackage{amssymb}

\usepackage{subfigure}
\usepackage{amsmath}

\usepackage{color}
\usepackage{ulem}

\journal{Journal of Theoretical Biology}

\begin{document}

\begin{frontmatter}
\title{Random walk in genome space: A key ingredient of intermittent dynamics of community assembly on evolutionary time scales}


\author[label1]{Yohsuke Murase}
\author[label1]{Takashi Shimada}
\author[label1]{Nobuyasu Ito}
\address[label1]{Department of Applied Physics, School of Engineering, The University of Tokyo, 
7-3-1 Hongo, Bunkyo-ku, Tokyo 113-8656, Japan}

\author[label2]{Per Arne Rikvold}
\address[label2]{Center for Materials Research and Technology and Department of Physics, 
Florida State University, Tallahassee, FL 32306-4350, USA}

\begin{abstract}
Community assembly is studied using individual-based multispecies models.
The models have stochastic population dynamics with mutation, migration, and extinction of species. 
Mutants appear as a result of mutation of the resident species, while migrants have no correlation with the resident species. 
It is found that the dynamics of community assembly with mutations are quite different from the case with migrations. 
In contrast to mutation models, which show intermittent dynamics of quasi-steady states interrupted by sudden reorganizations of the community, 
migration models show smooth and gradual renewal of the community. 
As a consequence, instead of the $1/f$ diversity fluctuations found for the mutation models, 
$1/f^2$, random-walk like fluctuations are observed for the migration models. 
In addition, a characteristic species-lifetime distribution is found: a power law that is cut off by a ``skewed'' distribution in the long-lifetime regime.
The latter has a longer tail than a simple exponential function, which indicates an age-dependent species-mortality function.
Since this characteristic profile has been observed, both in fossil data and in several other mathematical models,
we conclude that it is a universal feature of macroevolution.
\end{abstract}

\end{frontmatter}

\section{Introduction}\label{sec:introduction}

Understanding the dynamics of biological macroevolution has been one of the most challenging topics in evolutionary theory.
As more reliable and comprehensive fossil data are accumulated,
their statistical aspects have attracted increasing interest \citep{Alroy:2001uf,Newman:2001oq}.
Meanwhile, empirical studies of currently existing ecological communities provide snapshots of the evolutionary  dynamics.
Such fossil data and contemporary field data have previously tended to be discussed separately.
It has recently been recognized that issues of ecological and evolutional scales can in fact be strongly linked \citep{Thompson:1998bs,Thompson:1999fy,Yoshida:2003qy}. 
Consequently, several models have recently been proposed to bridge the gap between ecological and evolutional timescales. 
These include the tangled-nature model \citep{PhysRevE.66.011904,CHRISTENSEN:2002yq,0305-4470-36-4-302}, 
simplified versions of that model \citep{rikvold2003pea,zia-jpa,0305-4470-38-43-005,rikvold2007ibp,Rikvold:2007lr,Filotas:2010th,filotas-TaNaA}, 
the Webworld model \citep{Caldarelli:1998eu,drossel01:_influen_of_predat_prey_popul,Drossel:2004fj,McKane:2004qy}, 
and others \citep{PhysRevLett.90.068101,shimada-arob2002,tokita-tpb2003,rikv2009}.
These models consist of coupled population dynamics for each species, 
complemented by rules for introducing new species to the community.
Survivability of individuals (or species) are determined by the totality of extant species.
In other words, species undergo density dependent selection, which can also be seen as selection in a dynamically changing fitness landscape \citep{Gavrilets:1997fk,gavrilets:fla}.
This picture is in contrast with neutral models, in which no strong interspecies interactions are included \citep{hubbell2001unified,pigolotti2005sld}.

In this article, the relation between temporal patterns on evolutionary timescales and
community dynamics on finer ecological scales are studied with individual-based models.
Simple stochastic processes, such as random walks and branching processes, have often been used for the interpretation of fossil data.
However it is still an open question which aspects of evolution and extinction dynamics can be interpreted by such simple processes.
In communities with complex interactions, do species compositions change gradually or intermittently showing coordinated stasis \citep{DiMichele:2004kx}? 
What are the necessary conditions for the ``Red-Queen'' hypothesis \citep{van1973new,Raup:1975oq,DORAN:2006la,Finnegan:2008ts,Benton:2009qe} to be valid?
Do more realistic population-dynamics models yield large avalanches as predicted in simplistic SOC models \citep{bak93:_punct_equil_and_critic_in,newman2003me}?
Answering these questions by theoretical model development will contribute to the interpretation of the statistics of fossil records
and will also give insight into the conservation of currently existing communities.

Deliberately simple population-dynamics models with species turnover should answer questions about universal features of community assembly.
The models considered here are simplified versions of the tangled-nature models 
\citep{rikvold2003pea,rikvold2007ibp,Rikvold:2007lr}. 
The tangled-nature model is a simple individual-based model, originally introduced 
by Hall and co-workers \citep{PhysRevE.66.011904,CHRISTENSEN:2002yq} 
and later simplified by \citet{rikvold2003pea}. 
The evolutionary dynamics of various versions and derivatives of the tangled-nature model have been studied extensively.
These previous studies have revealed that the model communities evolve intermittently rather than gradually. 
The evolution process consists of quiet periods during which the species composition of the community remains nearly constant, reminiscent of coordinated stasis \citep{DiMichele:2004kx},
and brief, active periods during which drastic rearrangement of the community composition takes place.
The intermittency gives rise to approximate $1/f$ fluctuations of the species diversity, 
approximate $t^{-2}$ species-lifetime distributions, and 
power-law duration distributions for quasi-steady states (QSSs).
The $1/f$ fluctuations are distinct from those of the naive random walk: 
fluctuations on short and long timescales are self-similar and strongly correlated.
Interestingly, these statistical features are observed not only for mutualistic communities \citep{rikvold2003pea,Rikvold:2007lr}, 
but also for several predator-prey models \citep{rikvold2007ibp,Rikvold:2007lr,rikv2009}. 
Even though the network structures developed in these two types of 
communities are quite different \citep{Rikvold:2007lr}, these dynamical statistics are similar. 
Furthermore, neither model shows significant dependence on the choice of parameter sets \citep{Rikvold:2007lr}. 
This implies that the observed dynamical features are universal with respect to the form of population dynamics,
and supports the validity of such simple population-dynamics models.
We expect that the dynamics on evolutionary timescales are characterized by a few key factors, 
and here we are aiming to identify some of these.

In this article, we show that the way in which new species are introduced into the community plays a significant role for the stability of QSSs
by comparing two models, called ``mutation'' and ``migration.''
In the ``migration'' model, we consider community assembly via totally random migration of new species transported from another community.
Exotic immigrants are generally uncorrelated with native species since they have evolved in different environments. 
Therefore, any kind of new species may appear, regardless of the resident species. 
On the other hand, possible candidates for new species may often depend on the species composition of the resident community, 
for example in the case that the community is isolated and its evolutionary driving force is mainly sympatric speciation.
This aspect is modeled in the ``mutation'' model.
If the diversity of potential newcomers is limited, the community is expected to be more stable because it only has to be resilient  against a limited number of new species.
This limitation is expected to have a significant effect on the stability of the community and its dynamics on evolutionary timescales.
In nature, some exotic invasions cause losses in the biological diversity of native species, 
and they are a major cause of decline in global biodiversity \citep{Lodge1993133,Filotas:2010th,filotas-TaNaA}. 
Exotic invaders also have major economical consequences amounting to billions of dollars. 
Thus, this effect on ecosystems is becoming 
an important focus of ecological study \citep{Lodge1993133,carlton1993erg,Vitousek:1997vl,Kolar2001199,Carlton:1996hs}.

\section{Models}

Two forms of population dynamics (Model A and Model B) 
and two rules for the introductions of new species (migration and mutation) are investigated. 
In the following, we term the Model A (B) with the mutation (migration) rule ``mutation (migration) Model A (B).''

\subsection{Reproduction probability}\label{subsec:tn}

The models considered here are 
the simplified versions of the tangled-nature model \citep{rikvold2003pea,rikvold2007ibp,Rikvold:2007lr}. 
In these models, 
the population evolves stochastically in discrete, non-overlapping generations. 
Each individual of species $I$ gives rise to $F$ offspring 
with a reproduction probability $P_I$ before it dies. 
Otherwise it dies without offspring. 
The reproduction probability $P_I$ for an individual of species $I$ in generation $t$ 
depends on the individual's ability to utilize the amount $R$ of available external resources, 
and on its interactions with the population sizes $n_J(t)$ of all the species present 
in the community at that time.
The form of $P_I$ is taken as 
\begin{equation}\label{eq:p}
	P_I(R,\{n_J(t)\}) = \frac{1}{1+\exp{ \left[ - \Delta_I(R,\{n_J(t)\}) \right] } } \;,
\end{equation}
where
\begin{equation}\label{eq:delta}
	\Delta_I(R,\{n_J(t)\}) = - b_I + \frac{\eta_I R}{N_{\rm tot}(t)} 
	+ \sum_J \frac{M_{IJ} n_J(t)}{N_{\rm tot}(t)} - \frac{N_{\rm tot}}{N_{\rm 0}}.
\end{equation}
Here $b_I$ is the cost of reproduction for species $I$ (always positive), 
and $\eta_I$ is the ability of individuals of species $I$ to utilize the external resource $R$. 
The matrix $\mathbf{M}$ defines the interactions between species.
The total population size is $N_{\rm tot}(t) = \sum_J n_J(t)$, 
and $N_0$ is an environmental carrying capacity that prevents $N_{\rm tot}(t)$ from diverging to infinity.
The reproduction probability $P_I(R,\{n_J(t)\})$ is a monotonically increasing function of $\Delta_I$, 
ranging over $(0,1)$.
For a large positive $\Delta_I$ 
(small birth cost, strong coupling to the external resource, and more prey than predators), 
$P_I$ approaches unity and the population of species $I$ increases.
In the opposite limit of large negative $\Delta_I$ 
(large birth cost, weak or no coupling to the external resources, and/or more predators than prey),
$P_I$ goes to zero and the population decreases rapidly. 
We also note that the results shown below do not depend on the assumption that the number of offspring per individual is always $F$.
For a reasonable probability distribution of $F$, the results should be similar \citep{Murase:2010fk}.

Two types of reproduction probabilities are considered in this article: Model A and Model B.
Model A has no restriction on the form of the interaction matrix $\mathbf{M}$. 
Therefore, each species makes various types of interactions with others, 
including predator-prey, mutualistic, and competitive interactions.
Model B focuses on the energy transport through a food web,
so the off-diagonal part of $\mathbf{M}$ is limited to being antisymmetric ($M_{IJ} = -M_{JI}$).
Thus, if $M_{IJ} > 0$ and $M_{JI} < 0$, then species $I$ is the predator and $J$ the prey. 

In Model A, the reproduction cost $b_I$ and the external resource $R$ are zero; 
thus the first and the second terms of Eq.~(\ref{eq:delta}) disappear.
The total population size is limited by the last term, which includes the carrying capacity $N_{\rm 0}$.
The off-diagonal elements of the interaction matrix $M_{IJ}$ are randomly drawn 
from a uniform distribution over $[-1,+1]$, 
while the diagonal elements are set to zero.
For Model A, $F=4$ and $N_0 = 2000$ are used in this article.
This particular value of $F$ is chosen such that perturbations away from the fixed point in the single-species limit with vanishing mutation rate will decay monotonically,
without oscillations or chaotic behavior \citep{rikvold2003pea,Filotas:2010th,filotas-TaNaA}.
As shown in \citep{rikvold2003pea,zia-jpa,Rikvold:2007lr,Filotas:2010th,filotas-TaNaA}, communities tend to evolve toward mutualism in Model A.

In Model B, the external resource $R$ is introduced. 
All the species have positive values of the birth cost $b_I$, which are randomly drawn from $(0,1]$, 
and a certain proportion ($0.05$ is used in this article) of species can feed on the resource, 
i.e., the resource couplings $\eta_I$ are positive for primary producers or autotrophs, 
and zero for consumers or heterotrophs. 
The resource $R$ remains constant (here, $2000$).
The off-diagonal part of the interaction matrix is antisymmetric. 
Non-zero elements are assigned randomly to $M_{IJ}$ ($=-M_{JI}$) for $I < J$ with probability $c = 0.1$,
which is consistent with the connectance of food webs in nature \citep{Dunne:2002oq,Dunne:2002eu}. 
The nonzero elements of the interaction matrix are randomly chosen from a triangular distribution on $[-1,+1]$. 
The diagonal elements of $\mathbf{M}$, which represent the intraspecies interactions, 
are selected randomly from a uniform distribution on $[-1,0)$ for all the species. 
The environmental carrying capacity term is not included in this model ($N_0 = \infty$). 
The birth-cost term and the negative diagonal elements $M_{II}$ 
prevent species populations from growing to infinity. 
The fecundity $F$ for Model B is set to $2$ for the same reasons as discussed for Model A above.
These models have fixed-point populations $\left\{ n_{I}^{\ast} \right\}$, 
which can be calculated exactly in the absence of mutations and migrations \citep{rikvold2007ibp,Rikvold:2007lr}. 
The dynamics of the populations are asymptotically stable 
but $\{n_I\}$ fluctuate around their fixed points due 
to the demographic stochasticity \citep{zia-jpa}.\footnote{
Strictly speaking, the unique stationary state for the mutation models 
is the state in which all species are extinct. 
However, the fluctuations around the locally stable fixed point,  
needed to cause complete extinction, are so extreme that the lifetime of
a nonzero population size should be $O(e^{N_0} )$ 
or $O(e^{R})$, and thus for all practical purposes infinite \citep{zia-jpa}. 
}

These models are advantageous since the fixed point and the linear stability 
can be analytically estimated. 
For the sake of comparison we used the same parameter sets as in 
earlier articles on the same models,
but the results shown in the next section do not show qualitative differences for other reasonable parameter sets.

\subsection{Introduction of new species: migration versus mutation}\label{subsec:addition}
Introduction of new species and extinction of existing ones are essential for a macroevolution process of community assembly.
New species are added to the system by ``migration'' or ``mutation,'' 
which are different rules for the introduction of new species. 
A species whose population becomes zero goes extinct. 

In the ``migration'' models, an individual of a new species appears in the community with properties ($b_{I}$, $\eta_I$ and $M_{IJ}$) randomly assigned from the specified distributions.
Thus, there is no correlation between the species existing in the system and the immigrants. 
Once a species goes extinct, it never appears again. 
The migrations happen at regular intervals of length $\tau$.
This is definitely the simplest way to introduce new species into the system. 

In contrast to the ``migration'' model,
in the ``mutation'' model the candidates for new species depend on the resident species.
We consider the case that new species are limited to the ones that are phylogenetically close to the resident species. 
In order to include this effect, a coarse-grained ``genome'' space is introduced. 
Each individual has a bit-string ``genome'' of length $L$ \citep{PhysRevE.66.011904,CHRISTENSEN:2002yq,0305-4470-36-4-302,gavrilets:fla}.
Each bit sequence corresponds to a different species, which has a different phenotype from the neighboring species.
Thus, the total number of potential species is $2^{L}$.
All the values of $b_I$, $\eta_I$, and $M_{IJ}$ 
are predetermined for each genotype at the beginning of the simulation. 
In every generation, a mutation may happen to the genomes of the offspring:
all the genome bits of appearing offspring, which amount to $N_{\rm tot} \times L$, 
flip independently with a probability $\mu/L$, so that the average number of mutations per offspring individual is $\mu$. 
These flips result in the appearance of new species
and can be seen as a random walk along the edges of the hypercube defined by the states of the $L$-bit genome \citep{Gavrilets:1997fk,gavrilets:fla}.

The mutation rate per species, $\mu$, determines how frequently new species appear. 
Thus, an individual moves randomly to a neighboring site in the $L$-dimensional hyper-cubic space by a mutation.
The probability of $m$-bit mutations in a single individual is small, $O(\mu^m)$.
The properties of species $I$ ($b_I$, $\eta_I$, and $M_{IJ}$) have no correlation with those of its neighbor species.
This is a clearly idealized aspect of the model since phylogenetically related species usually have phenotypic correlations.
However, this model works as a good starting point for comparison with the migration model because we can focus on the effects of the phylogenetic correlation. 
In particular we note that the dynamic insensitivity of mutation Model A to correlation between the traits of parents and mutants was demonstrated by \citet{0305-4470-38-43-005}.

Although the mutation model can be considered as a quasi-species model as well \citep{Eigen:1977yf,0305-4470-36-4-302}, 
the mutation used here does not necessarily refer to a point mutation of an actual genome when the model is discussed in the context of community assembly.
The genome space is introduced in order to express the phylogenetic distance between two species and 
a bit-flip of the coarse-grained genome may correspond to a big leap in an actual genome space.
The mutation model fits better than the migration model for cases 
where new species are always phylogenetically close to the resident species.
Communities in isolated islands or lakes where sympatric speciation is the major driving force of evolution may be relevant examples.
On the other hand, communities geographically neighboring to a large species pool, such as communities in a continent,
may be better described by the migration model.
A detailed description of the speciation process is not included in either model.

The crucial difference between the migration and mutation models is that the mutation model limits the variety of new species accessible from a given community. 
The reason why we use a binary string instead of a set of real numbers is 
that a bit-string maximizes the potential number of species while keeping the number of neighbor species reasonably small.
Models with more alleles would be expected to show intermediate properties between the results for the migration and mutation models.
Although it is hard to estimate how many neighboring species should be assumed, 
the comparison of these two extreme models highlights the effect of this limitation.
Before adopting more complex models, we therefore start from these simple cases.

\section{Results}\label{sec:results}
\subsection{Typical time series}
First we show typical time series of kinetic Monte Carlo simulations for the four models. 
Genome length $L = 25$ and $22$ are used for mutation Model A and B, respectively.
Therefore, the numbers of potential species are $2^{25}$ and $2^{22}$, respectively.
The results shown in the following do not depend significantly on the precise values of $L$.
The mutation rate $\mu = 0.001$ is used for the mutation models, 
and the migration interval $\tau = 1$ is used for the migration models.
This mutation rate is selected for computational feasibility although it might be rather high compared with actual ecosystems. 
We confirmed that the power law behaviors shown below remain similar for mutation rates as small as $10^{-4}$, and intermittency was qualitatively observed for $\mu = 10^{-6}$.
The dynamic robustness of of mutation models A and B with respect to the model parameters ($L$, $\mu$, $N_0$, $R$) is discussed in Appendix C of \citet{Rikvold:2007lr}.
In particular, it is shown there that the diversities depend only sublinearly on Hubbell's fundamental biodiversity number, $2R \mu$ or $2N_0 \mu$,
which is proportional to the number of mutant organisms produced in each generation \citep{hubbell2001unified}.
It is therefore justified to use large mutation rates in combination with small population sizes to overcome computational limitations. 

In this article, 
the exponential Shannon-Wiener diversity index \citep{Krebs:1989pi} is used as a measure of biodiversity. 
This index is defined 
as the exponential function of the information entropy of the population distribution, 
$D(t) = \exp [S\left(\{n_I(t)\}\right)]$, 
where $S\left( \left\{ n_I\left(t\right) \right\} \right) = -\sum_{ \left\{I|\rho_I(t)>0\right\} } \rho_I(t)\ln{ \rho_I(t) }$,
with $\rho_I(t) = n_I(t)/N_{\rm tot}(t)$. 
We adopted this measure in order to filter out unsuccessful mutants or migrants 
which have tiny populations and rapidly go extinct. 

\begin{figure}[!ht]
\begin{center}
\includegraphics[width=.5\textwidth]{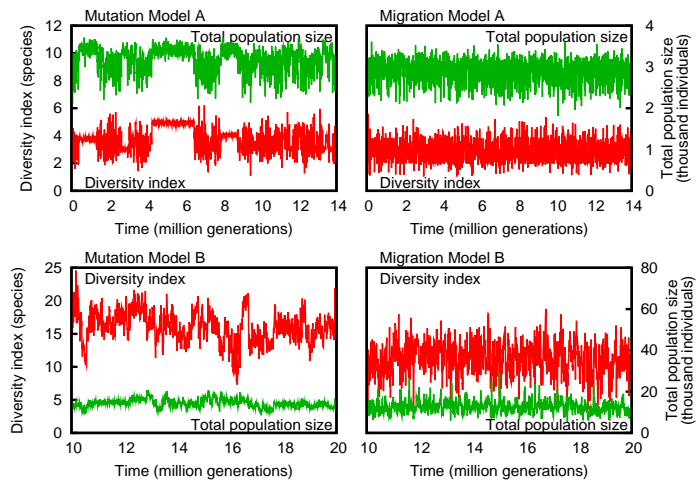}
\caption{
(Color online) Typical time series of exponential Shannon-Wiener diversity index and total  population size 
for mutation and migration Models A and B. 
Mutation rate $\mu = 0.001$ is used for the mutation models, and migration interval $\tau = 1$ is used for the migration models.
Genome length $L$ for mutation Model A and B are $25$ and $22$, respectively.
\label{fig:timeseries}}
\end{center}
\end{figure}

Figure~\ref{fig:timeseries} 
shows the dynamics of the diversity indices 
and the total population size for typical simulation runs for each of the four models. 
The mutation models show intermittent behaviors, consisting of active and quiet periods. 
During the active periods, the diversity and the total population size show larger fluctuations, 
and the species composition changes quickly. 
During the quiet periods, the species composition remains nearly constant, 
and the system is considered to be in a quasi-steady state (QSS). 
The community assembly proceeds intermittently, rather than gradually.

In contrast, the intermittent behaviors are hardly observable in the migration models. 
It is difficult to find QSSs, at least from these figures.
The system always fluctuates actively, indicating continuous renewal of the species composition. 
Thus, the assembly dynamics for the migration models are significantly different from the mutation models. 
A key ingredient of the intermittency in the evolution dynamics is the introduction of the genome space, 
which limits the number of different mutants that can be produced by a given community.

\subsection{Statistics of dynamical behaviors}
To evaluate the intermittency quantitatively, 
we calculated several statistics of the dynamical behaviors for each model.
We performed simulations of $2^{25}$ generations with 
$2^{22}$ generations as a ``warm-up'' period for Model A, 
and simulations of $2^{26}$ generations with $2^{24}$ generations warm-up for Model B.
These warm-up periods are long enough to realize statistically stationary 
states. 
For each model and species-introduction mechanism, the data were averaged 
over six independent runs. 
Each simulation run was started with a single, randomly chosen species 
(producer species for Model B) with a population size of $100$ individuals. 
The details of this initial condition are totally insignificant, 
and the systems were completely ``thermalized'' 
during the initial warm-up periods.

\begin{figure}[!ht]
\begin{center}
\subfigure{
\includegraphics[width=.45\textwidth]{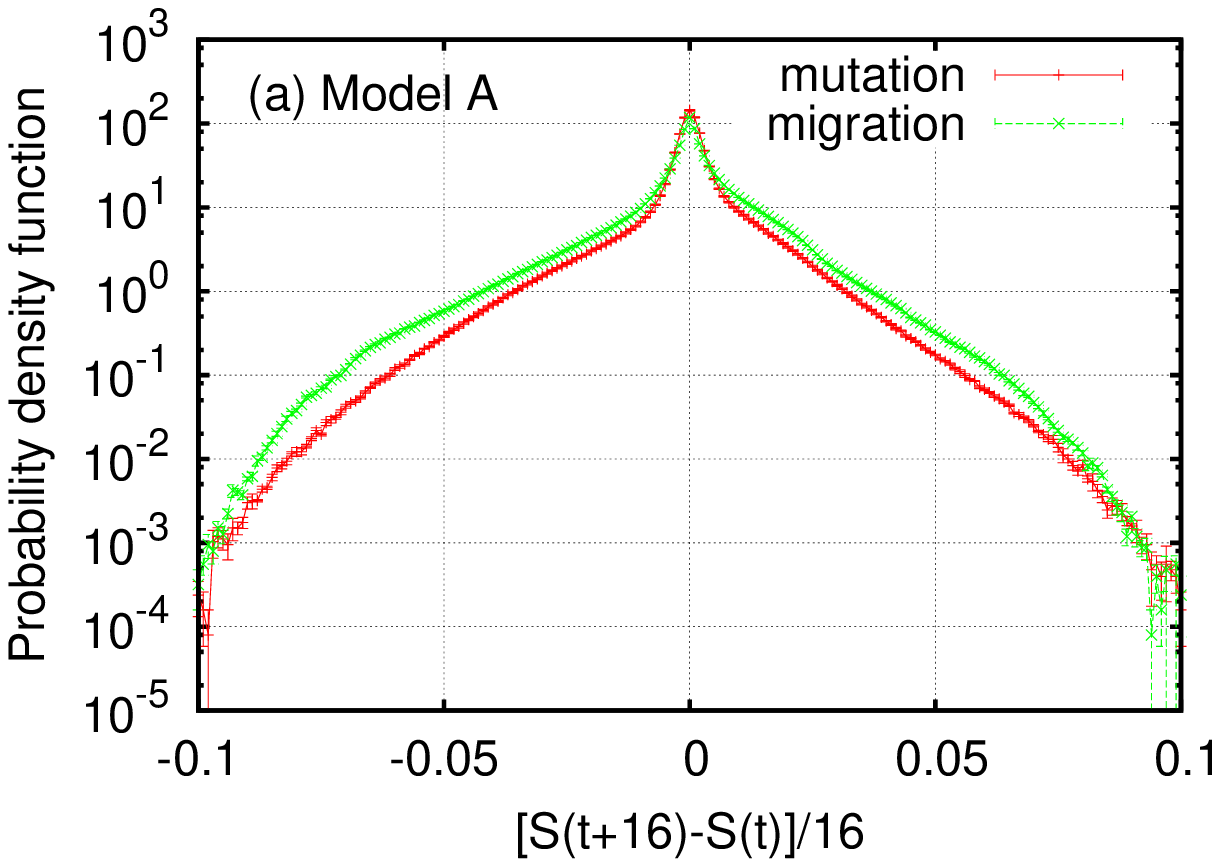}
\label{fig:a-dsdt}
}
\subfigure{
\includegraphics[width=.45\textwidth]{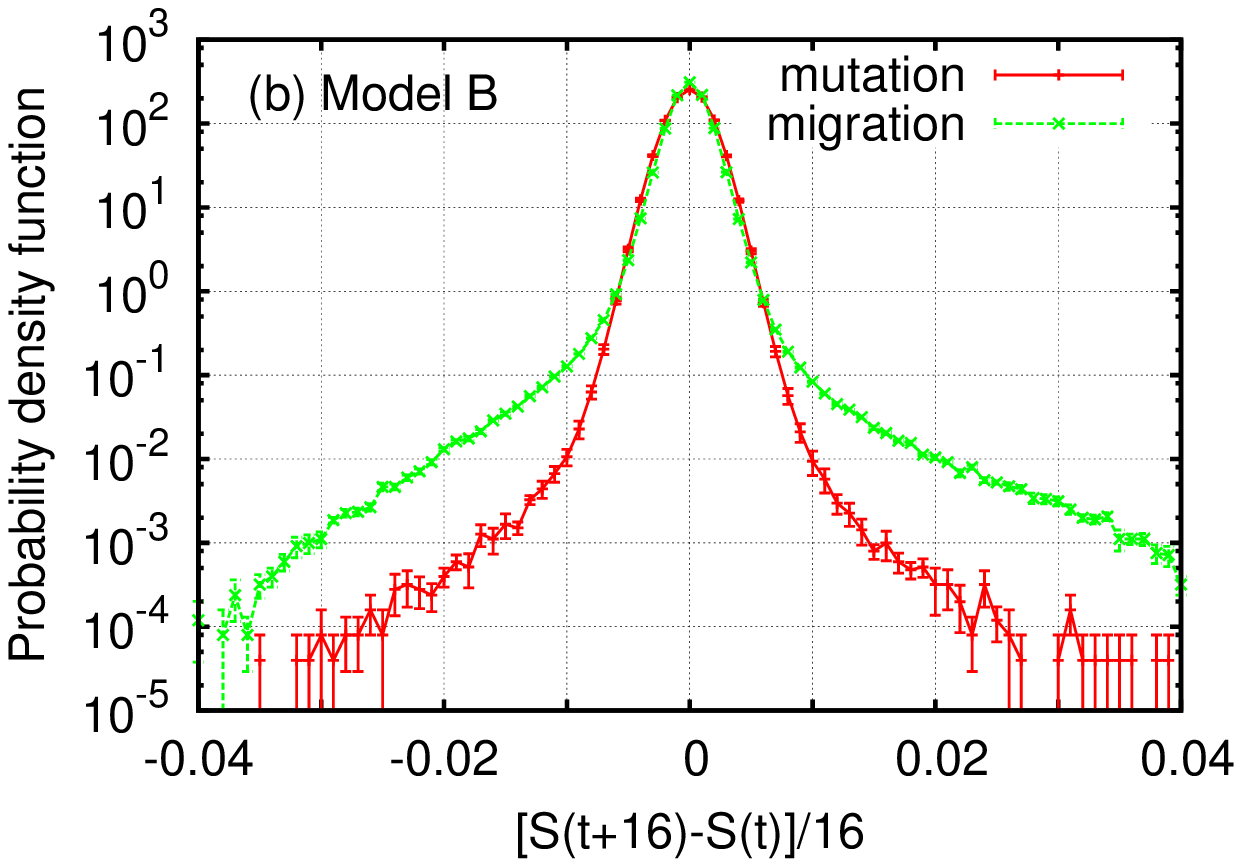}
\label{fig:b-dsdt}
}
\caption{
(Color online) Normalized histograms of the logarithmic derivative of the diversity index for Model A (a) and Model B (b).
Mutation rate $\mu = 0.001$ is used for the mutation models, and migration interval $\tau = 1$ is used for the migration models.
\label{fig:dsdt}}
\end{center}
\end{figure}

First, we estimated the QSS durations and calculated their probability density functions (pdf). 
One way to identify QSS is to introduce a cutoff on the logarithmic derivative of the diversity, $dS/dt$. 
The pdf's of this quantity (averaged over $16$ generations) 
are shown in Fig.~\ref{fig:dsdt}. 
For all the models, each distribution has a sharp peak around the center 
and relatively wide wings in both tails \citep{rikvold2003pea,Rikvold:2007lr}. 
The sharp peak around zero represents that the community is in a quiet period, 
while the large wings represent large rearrangements of the species composition. 
QSSs are estimated as the periods between times when $|dS/dt|$ exceeds a cutoff. 
We adopted $0.015$ and $0.01$ as the cutoff for Model A and Model B, respectively. 
We note that the migration models have larger wings than the corresponding mutation models. 
This indicates that the migration models spend more time in active periods. 

\begin{figure}[!ht]
\begin{center}
\subfigure{
\includegraphics[width=.45\textwidth]{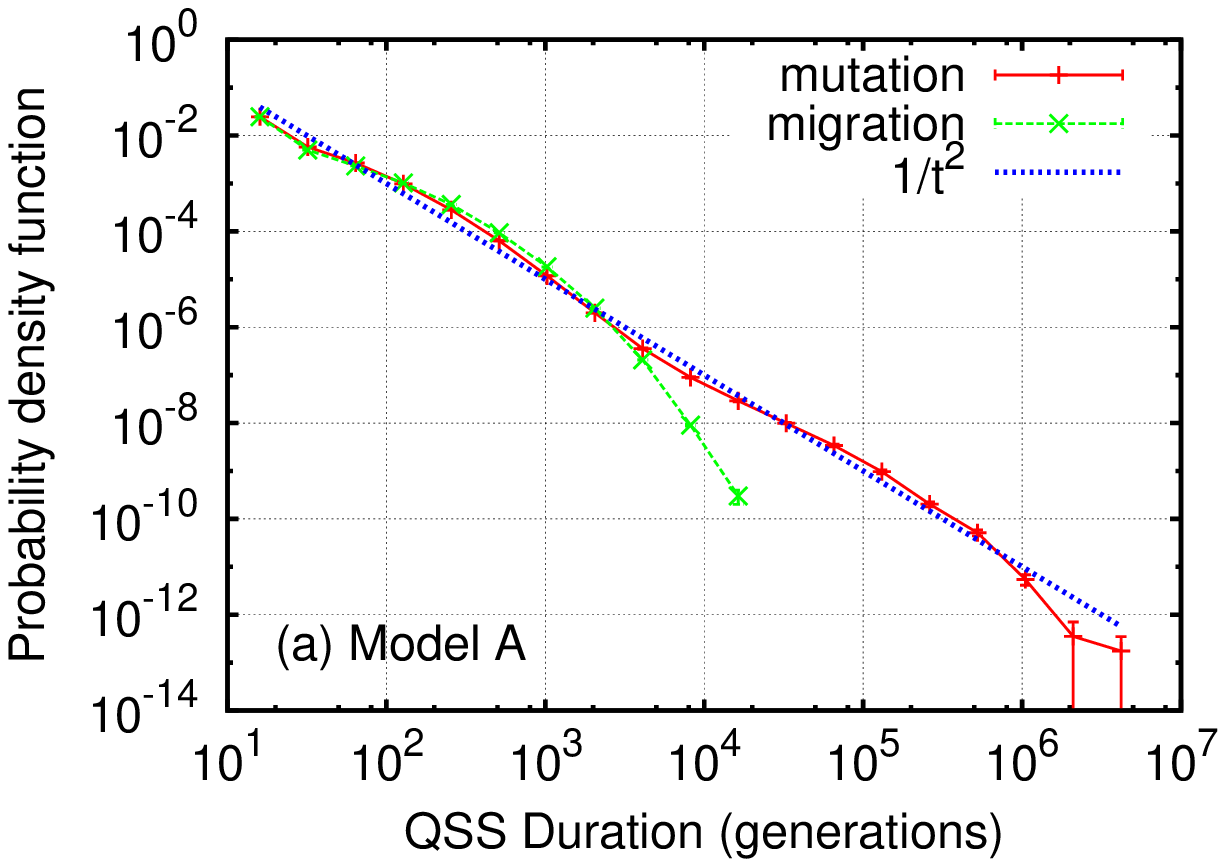}
\label{fig:a-duration}
}
\subfigure{
\includegraphics[width=.45\textwidth]{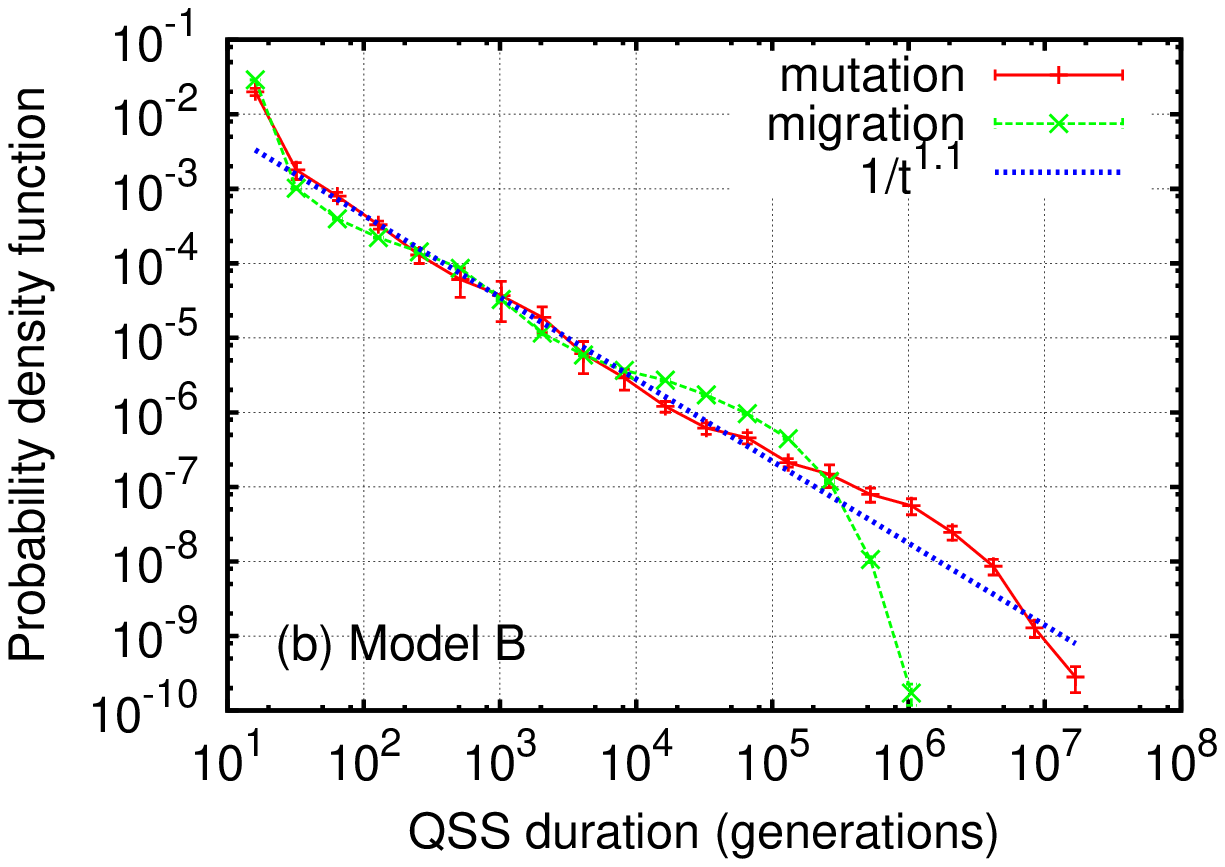}
\label{fig:b-duration}
}
\caption{
(Color online) Normalized histograms of the QSS durations for Model A (a) and Model B (b). 
Mutation rate $\mu = 0.001$ is used for the mutation models, and migration interval $\tau = 1$ is used for the migration models.
The thresholds for estimating the QSS durations are $0.015$ and $0.01$ for Model A and Model B, respectively.
\label{fig:duration}}
\end{center}
\end{figure}

The pdf's of the QSS durations are shown in Fig.~\ref{fig:duration}. 
The QSS duration distributions for the mutation models show approximate power laws.
Approximate $t^{-2}$ and $t^{-1}$ behaviors are observed up to $10^{7}$ generations for mutation Model A and B, respectively. 
Thus, the community-assembly dynamics for the mutation models have long-time correlations. 
On the other hand, the migration models show faster decays 
and deviations from the power laws at certain characteristic time scales. 
For migration Model A and B, it starts to decay faster than a power law at about $10^{3}$ and $10^{4}$ generations, respectively. 
While the distributions are well approximated by power laws up to the characteristic time scales,
another trend, which appears concave in a log-log plot, starts to emerge at longer time scales. 
Thus, the QSS distributions for the migration models have characteristic time scales, 
above which the distributions decay faster than power laws.
This concave curve is not fitted well by a simple exponential function, neither for migration Model A nor B. 
We discuss a fitting function for this concave curve, known as a $q$-exponential, in the Appendix.

\begin{figure}[!ht]
\begin{center}
\subfigure{
\includegraphics[width=.45\textwidth]{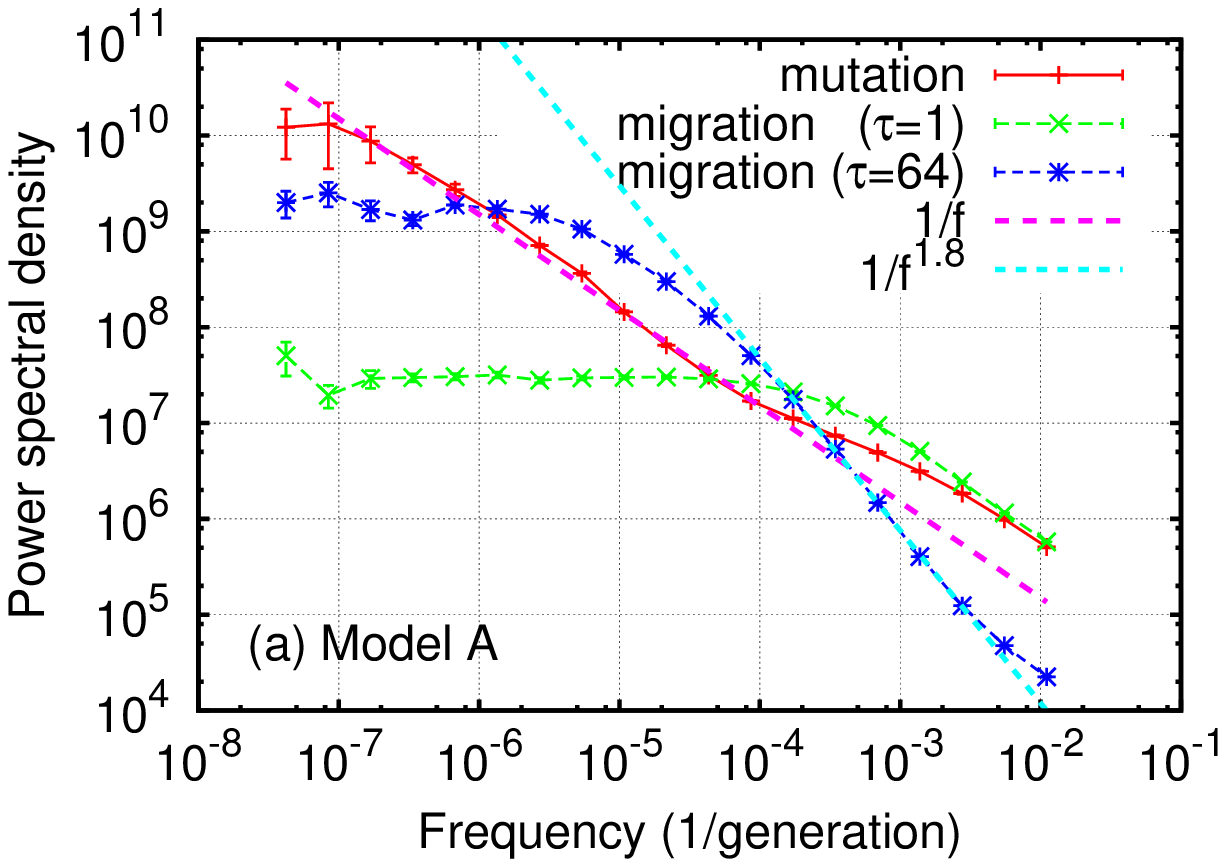}
\label{fig:a-psds}
}
\subfigure{
\includegraphics[width=.45\textwidth]{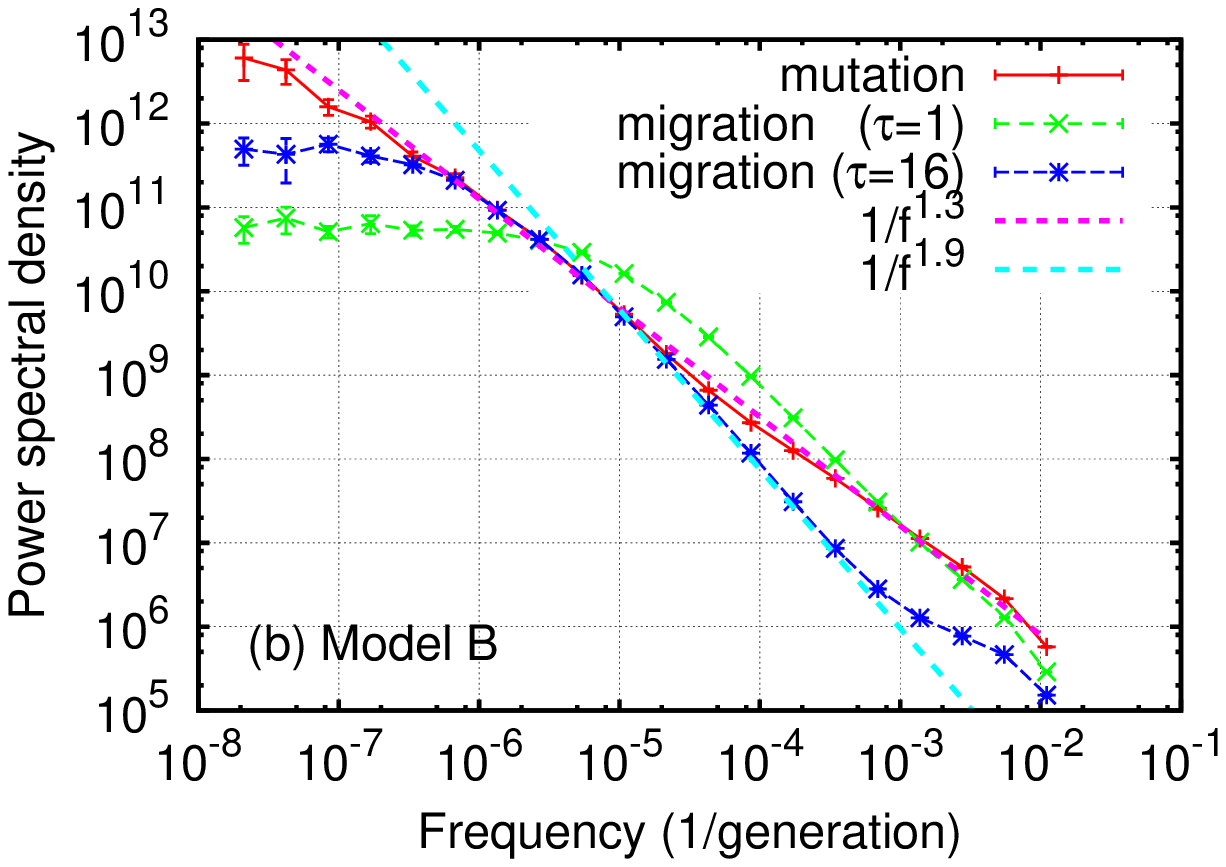}
\label{fig:b-psds}
}
\caption{
(Color online) Power spectral density of the diversity index for Model A (a) and Model B (b). 
Mutation rate $\mu = 0.001$ is used for the mutation models, and several values of $\tau$ are used for the migration models. 
\label{fig:psds}}
\end{center}
\end{figure}

The shorter QSSs for the migration models are also observed in power spectral densities (PSDs) of diversity indices. 
PSDs calculated for each model are shown in Fig.~\ref{fig:psds}.
For the mutation models, approximate $1/f$ fluctuations (flicker noise) are observed over several decades. 
This indicates that the evolution dynamics for the mutation models have quite long time correlations. 
On the other hand, for the migration models, the PSDs saturate at low frequency. 
This indicates that the temporal behavior of the diversity is uncorrelated for $t > 1/ 2\pi f_c$, 
where $f_c$ is the characteristic frequency at which the PSDs saturate. 
At $f > f_c$, the PSDs show approximate power laws $1/f^{\alpha}$. 
The exponent $\alpha$ for the migration models are larger than for the mutation models, 
and are close to $2$. 
Thus the dynamical behavior for the migration models are analogous to an Ornstein-Uhlenbeck process \citep{cox1977theory}.
in which the diversity index performs a random walk in an attractive ``potential.''
The optimum of this potential corresponds to a balance between extinctions and the invasion of new species.
The characteristic frequency for migration models is approximately proportional to $\tau^{-1}$. 
The exponent $\alpha$ for small $\tau$ is smaller than $2$, 
but this may be due to a transient from $1/f^2$ behavior to the saturation. 
We also calculated the PSDs of the total population sizes (not shown) and found the behaviors similar to the diversity indices.

\begin{figure}[!ht]
\begin{center}
\subfigure{
\includegraphics[width=.45\textwidth]{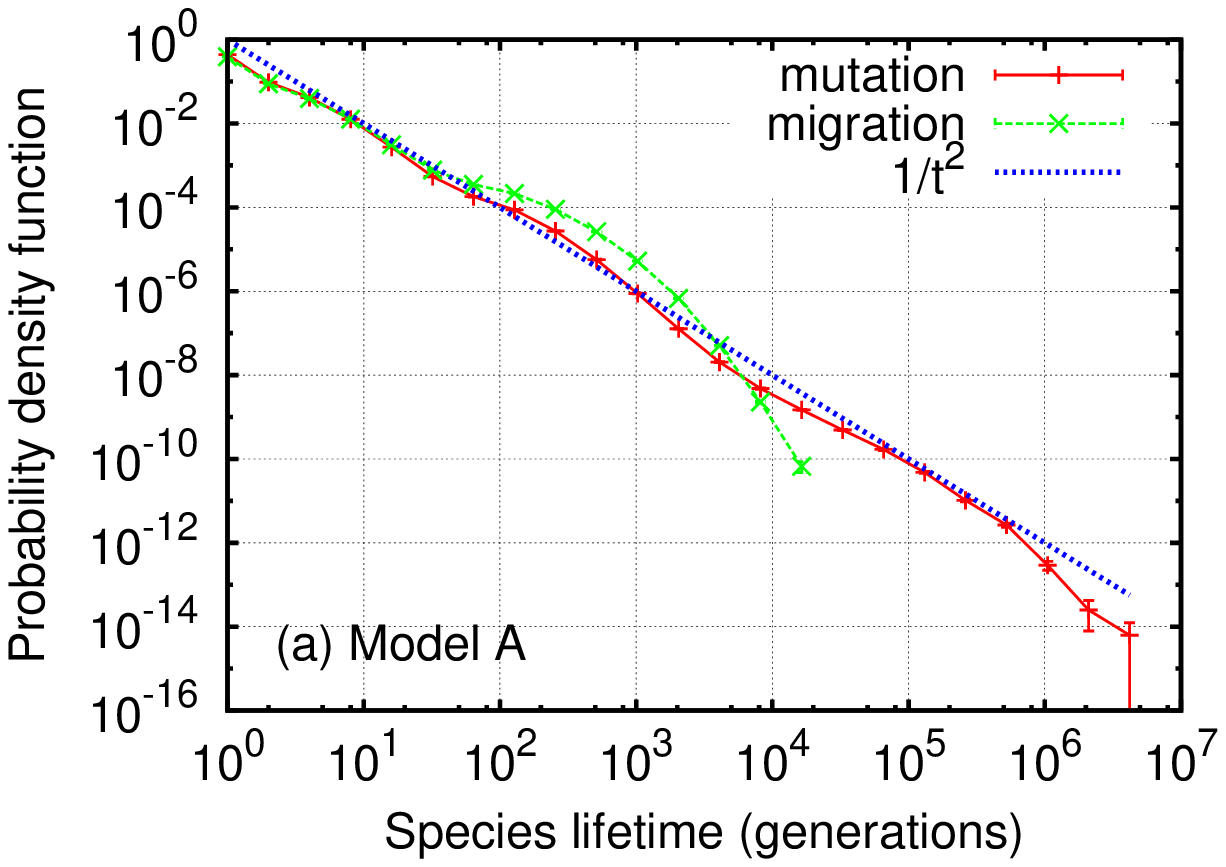}
\label{fig:a-lifetime}
}
\subfigure{
\includegraphics[width=.45\textwidth]{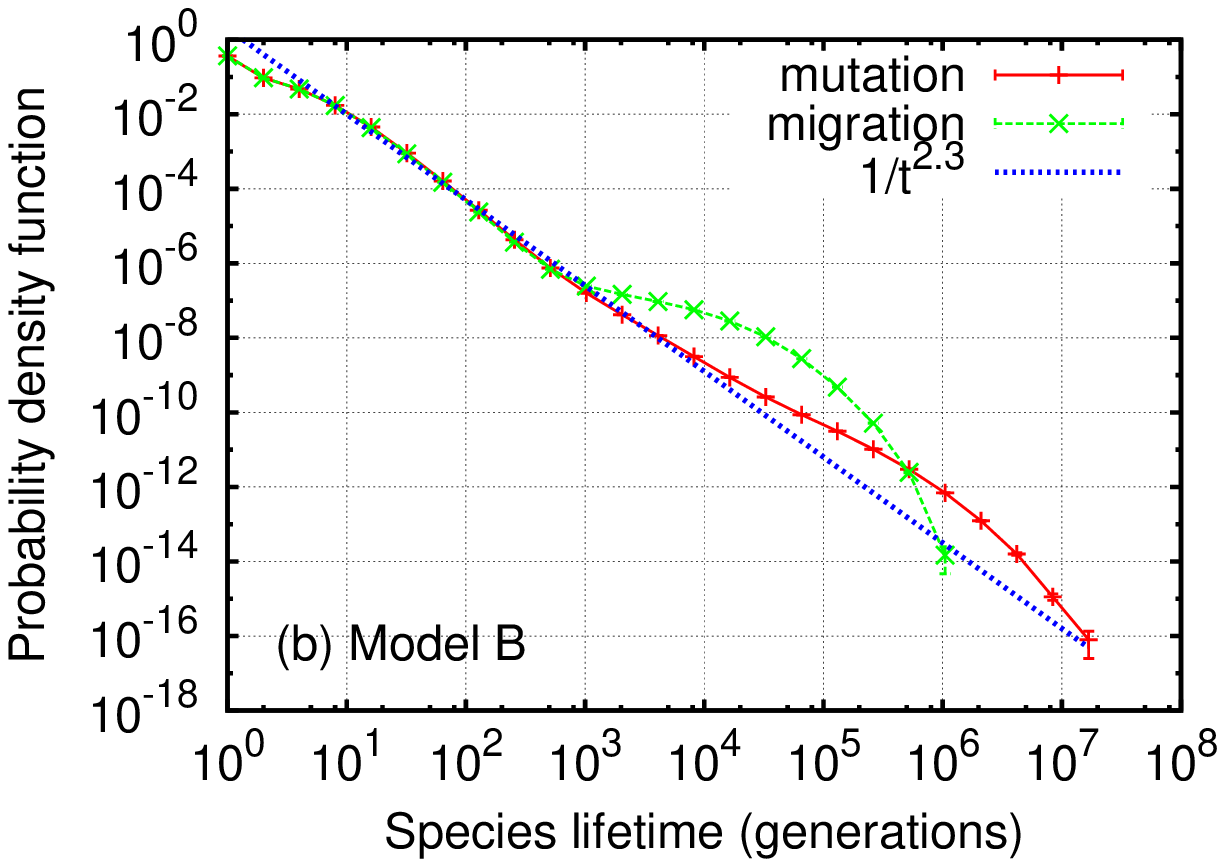}
\label{fig:b-lifetime}
}
\caption{
(Color online) Species-lifetime distributions for Models A (a) and Models B (b). 
Mutation rate $\mu = 0.001$ is used for the mutation models, and migration interval $\tau = 1$ is used for the migration models.
\label{fig:lifetime}}
\end{center}
\end{figure}

Figure~\ref{fig:lifetime} 
shows pdf's of the species lifetime for the four models. 
Both mutation Model A and B show approximate $t^{-2}$ power laws 
that continue up to around $10^7$ generations. 
Thus, species that live quite long appear relatively frequently for the mutation models. 
However, the migration models show a different profile. 
Although the distributions show power-law decays at small $t$, they cross over to skewed profiles (Appendix \ref{sec:tau-dep}), 
which look similar to those we see in the QSS duration distributions, 
i.e., concave on a log-log scale and convex on a semi-log scale. 
The shorter species lifetimes for the migration models are consistent with the shorter QSS durations 
since the rearrangement of communities cause large extinctions of species. 

We note clear similarities between the distributions of the QSS durations and of the species lifetimes for Model A.
It implies that the lifetimes of long-living species are mainly determined by the end of the QSS in which they live. 
In contrast, for Model B, the species-lifetime distributions are narrower than the QSS duration distributions.
It means extinctions of long-living species frequently happen during each QSS.

In conclusion, the migration models have a characteristic time scale in their dynamical behaviors, 
while the mutation models show power-law statistics. 
The differences between mutation and migration models are summarized in Table \ref{tab:summary}. 

\begin{table*}
\caption{Summary of the results for the four models studied. 
Representative behavior of QSS duration distributions (QSS), 
PSDs of the diversity and total population sizes (PSD), 
and species-lifetime distributions (SLD). 
} \label{tab:summary}
\begin{tabular*}{\hsize}{@{\extracolsep{\fill}}ccc} 
\hline
& mutation & migration \\
\hline
QSS & $1/t^2$ (Model A), $1/t$ (Model B) & $1/t^2$ ($1/t$) + skewed profile \\
PSD & $1/f$ & $1/f^2$ which saturates at low frequency \\
SLD & $1/t^2$ & $1/t^2$ + skewed profile \\
\hline
\end{tabular*} 
\end{table*}

\section{Discussion}\label{sec:discussion}

Several studies of plant and animal assemblages from fossil records have revealed long-term 
persistence punctuated by periods of rapid change \citep{DiMichele:2004kx,brett1995coordinated,brett1997coordinated}. 
Such intermittent patterns of community assemblages, termed coordinated stasis, have been found on the basis of fossil records although counterexamples are also common.
The possible origins of such intermittent patterns are still an open question.
One of the aims of the present study is to find conditions under which coordinated stasis can arise in models of macroevolution.

In previous studies of tangled-nature models with mutation 
\citep{PhysRevE.66.011904,CHRISTENSEN:2002yq,0305-4470-36-4-302,rikvold2003pea,rikvold2007ibp,Rikvold:2007lr}, 
it was found that the evolution proceeds intermittently rather than gradually. 
The systems spend most of their time in QSSs in which the species composition does not change significantly. 
The QSSs are interrupted by brief active periods, during which relatively large-scale community rearrangements happen. 
Duration distributions for QSS communities and species-lifetime distributions both show power laws. 

However, such QSSs become shorter and less well defined for migration models. 
The community-assembly dynamics is well described by an Ornstein-Uhlenbeck process, 
which is qualitatively different as it has a characteristic timescale.
Thus it appears that the introduction of a genome space may play a significant role for the sustainment of QSSs. 
It is remarkable that mutation considerably differs from migration, 
even when the elements of the interaction matrix for the mutation models have no correlations with each other. 
The biggest difference between the mutation and migration models is the variety of possible new species. 
For the mutation model, the number of possible mutants is roughly proportional to $L \times D^{\ast}$, 
where $D^{\ast}$ is the number of major resident species. 
If most of the candidates are not suited to survive in the existing QSS community, 
the species composition does not change, and the emergence and rapid extinction of unsuccessful mutants is just repeated. 
In this case, a successful species can emerge by a two-bit mutation of a large-population species 
or by a single mutation of a low-population species, both of which happen with very low probability. 
Therefore, species which are neighbors in genome space of resident species act as a ``protection zone.'' 
Intermittent dynamics in the community assembly can arise from the resulting limitation on the diversity of mutants.

Intermittency has often been discussed for complex population dynamics of 
a fixed set of species \citep{Gavrilets:1995wq,ives-ecology1998,Huisman-eco2001,Ferriere-ecol1999}. 
We note that the intermittency observed in the present study is different from those studies since the population dynamics itself is asymptotically stable. 
The intermittency is observed at the level of the rearrangement dynamics of community compositions, not at the level of the population dynamics.

We believe that the behaviors for the mutation models gradually become similar to the migration models 
when the genome length $L$ or the number of alleles increases. 
This is because the communities then are expected to have more paths to escape from the QSSs.
For isolated environments such as lakes or islands located far from another community, 
the mutation models would fit better, 
while the migration models would fit better for communities adjacent to large species pools.

Both for the migration and the mutation models, 
the species-lifetime distributions have heavier tails than simple exponential functions.
Thus, in communities with complex interspecies interactions, a naive Red-Queen hypothesis is not expected to hold.
The lifetime distribution estimated from the fossil data also shows a heavier tail than a simple exponential \citep{Finnegan:2008ts,Shimada:2003kx}.
(See also Fig.~\ref{fig:lifetime-fossil}.)
This fact indicates that extinctions due to the species interactions may explain the statistics in fossil records.

The models we studied show a fundamental difference from the SOC models.
Although large scale rearrangements of communities are observed (Fig.~\ref{fig:dsdt}),
the extinction-size distributions do not show power laws.
Therefore, these avalanches of extinctions are not critical processeses.
Thus, we speculate that
the key mechanism of the power laws observed for the mutation models are different from those of SOC models.

In a limit that all interactions are equal to zero and all $b_I$ (and $\eta_I$) are equal,
the models become similar to neutral models \citep{Rikvold:2007lr}.
In this limit, the species dynamics are characteristic of an off-critical branching process.
We have preliminary results for the neutral versions of the models.
The species-lifetime distributions show a $t^{-2}$ power law with an exponential cutoff,
which agrees with the return-time distribution of an off-critical branching process \citep{pigolotti2005sld}.
Intermittency is not observed for these models since no collective behavior between species is included.

Since a large range of randomly distributed interactions and species-specific parameters
(like $\eta_I$ and $b_I$) are available,
the population dynamics selects those combinations that 
give rise to quasi-steady states (i.e., quasi-stable communities). 
In model A, this yields compact, mutualistic communities.
The restriction to antisymmetric interactions in Model B limits the possibilities to predator-prey communities. 
The nonlinear character of the reproduction rate in our models make them more flexible to 
portray a wide variety of different communities and their dynamics simply through the 
evolutionary selection of realized parameters, than is the case for traditional Lotka-Volterra models.

\section{Summary}\label{sec:summary}

Four types of biological community-assembly models were studied. 
By comparison between mutation and migration, 
we clarified that 
the limitations on the variety of possible mutants resulting from the introduction of a genome space plays a key role 
in the emergence of QSSs and intermittent evolution dynamics that show power-law statistics over several decades.
This also indicates that exotic migrants can destroy QSSs more easily than mutants of resident species. 

The species-lifetime distributions for the migration models show a robust pattern: 
an approximate $t^{-2}$ power law, 
consistent with a stochastic branching process, 
which crosses over to a skewed profile. 
This skewed profile can be reasonably fitted by a $q$-exponential function, 
which is derived from an age-dependent species-mortality function. 
We believe that this skewed profile can be a natural candidate for the fitting of species-lifetime distributions 
in addition to simple exponentials and simple power-law functions.

We emphasize the robustness of our results. 
Even though Model A and B have significantly 
different types of interactions and as a result develop different network structures \citep{Rikvold:2007lr},
significant similarities are found between the models. 
Furthermore, the profile of species-lifetime distributions is similar to 
other models with quite different population dynamics \citep{PhysRevLett.90.068101,shimada-arob2002,Laird:2006it}. 
Further studies on both realistic and simplistic forms of population dynamics, such as the Webworld model, are planned.

\section*{Acknowledgments}
We are grateful for helpful comments on the manuscript
by A.~G.~Rossberg, V.~Sevim, and E.~Filotas, and for a useful conversation with L. H. Liow. 
This work was partly supported by 21st Century COE Program ``Applied Physics on Strong Correlation'' 
from the Ministry of Education, Culture, Sports, Science, and Technology of Japan, the JSPS (No. 19340110), 
and GRP of KAUST (KUK-I1-005-04), 
and a Grant-in-Aid for Young Scientists (B) No. 21740284 to T.~S.
from the Ministry of Education, Culture, Sports, Science and Technology of Japan.
Y.~M. appreciates hospitality at Florida State University, 
where work was supported by U.S. NSF Grant Nos. DMR-0444051 and DMR-0802288.

\appendix
\section{Fitting with a $q$-exponential function}\label{sec:tau-dep}

\begin{figure}[!ht]
\begin{center}
\subfigure{
\includegraphics[width=.45\textwidth]{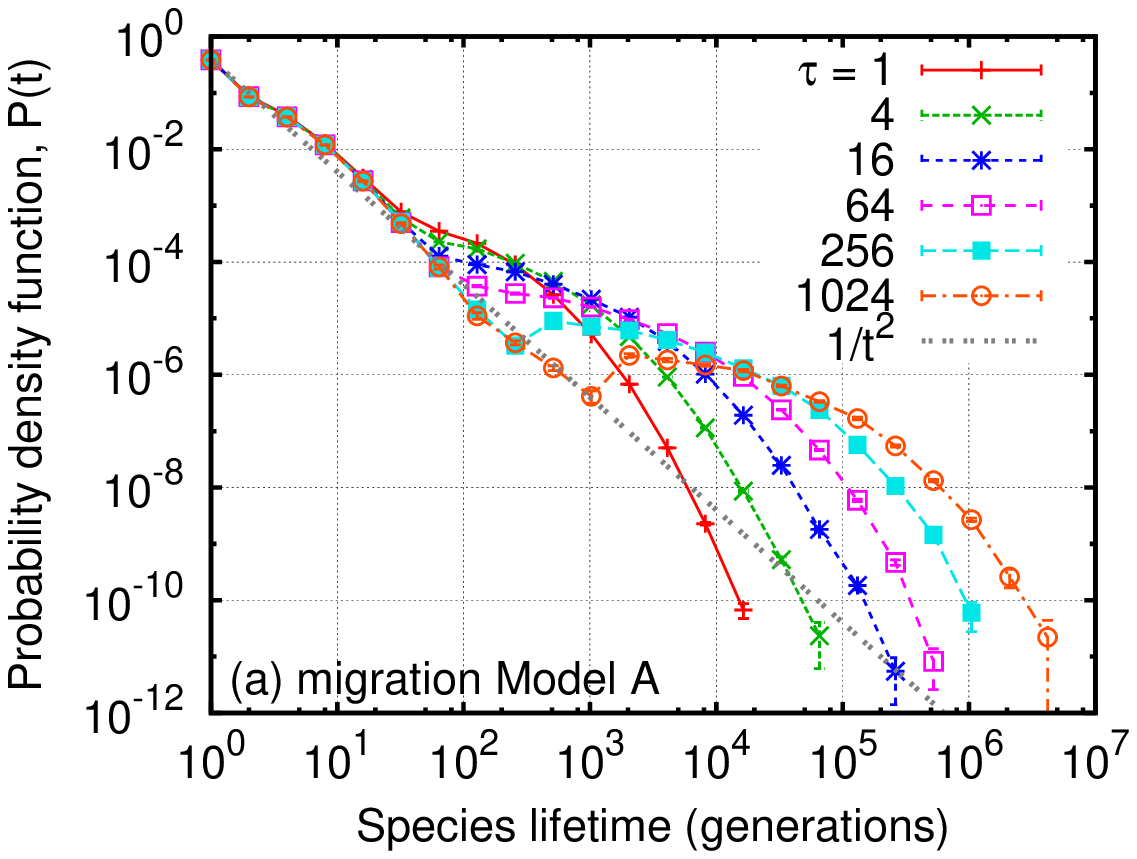}
\label{fig:a-inv-lifetime-taudep}
}
\subfigure{
\includegraphics[width=.45\textwidth]{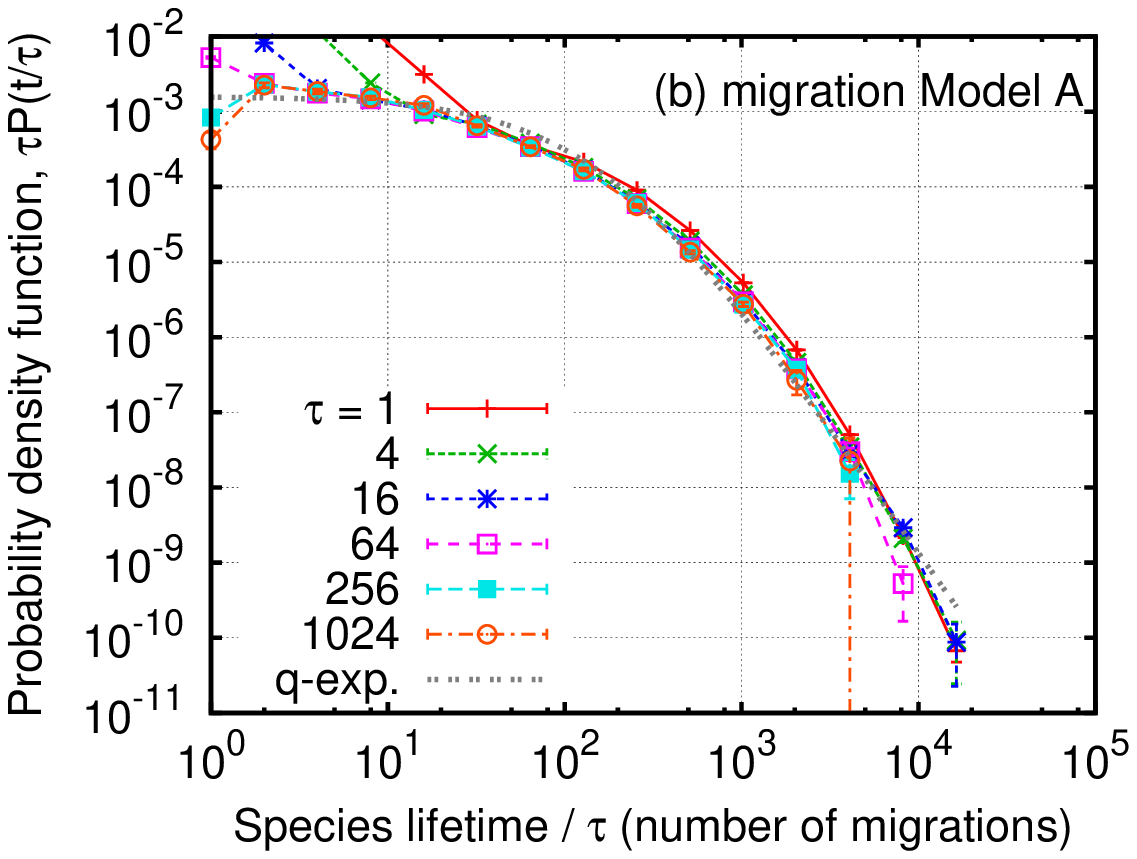}
\label{fig:a-inv-lifetime-normedtau}
}
\subfigure{
\includegraphics[width=.45\textwidth]{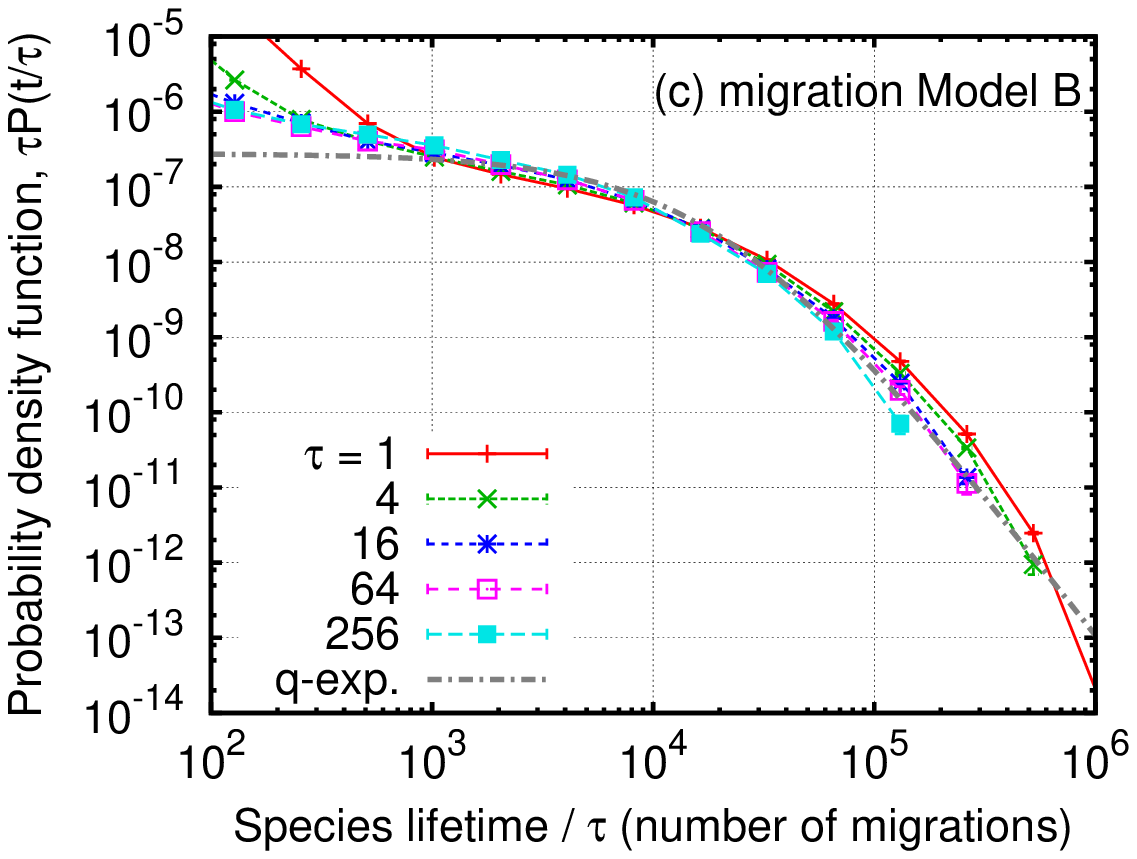}
\label{fig:b-inv-lifetime-normedtau}
}
\caption{
(Color online) (a) Species-lifetime distributions $P(t)$ for migration Model A with several values of $\tau$. 
(b) Species-lifetime distributions normalized by $\tau$, $\tau P(t/\tau)$, for migration Model A. 
It is fitted by a $q$-exponential function with $q=1.296$ and $\beta=0.0208$.
(c) Species-lifetime distributions normalized by $\tau$, $\tau P(t/\tau)$, for migration Model B, 
fitted by a $q$-exponential function with $q=1.26$ and $\beta=0.00018$.
\label{fig:lifetime-taudep}}
\end{center}
\end{figure}

Since the migration models have characteristic time scales in their community-assembly dynamics, 
we expect that these are related to the migration interval $\tau$.
Species-lifetime distributions for migration Model A with several values of $\tau$ are shown in Fig.~\ref{fig:a-inv-lifetime-taudep}. 
The skewed profile in the longer-time regime shifts to the right as $\tau$ increases, 
although the initial $t^{-2}$ power law does not show significant dependence on $\tau$. 
The amount of the shift is approximately proportional to $\tau$.
Thus, the distributions reasonably collapse onto a single curve for several values of $\tau$ 
by the rescaling of time by $\tau$ (Fig.~\ref{fig:a-inv-lifetime-normedtau}). 
Therefore, the typical time scale is determined by the number of migrations.

The same plot for migration Model B is shown in Fig.~\ref{fig:b-inv-lifetime-normedtau}. 
The data scales reasonably with $\tau$ although there are small deviations. 
We speculate that these deviations are caused by the stochastic population fluctuations.

\begin{figure}[!ht]
\begin{center}
\subfigure{
\includegraphics[width=.45\textwidth]{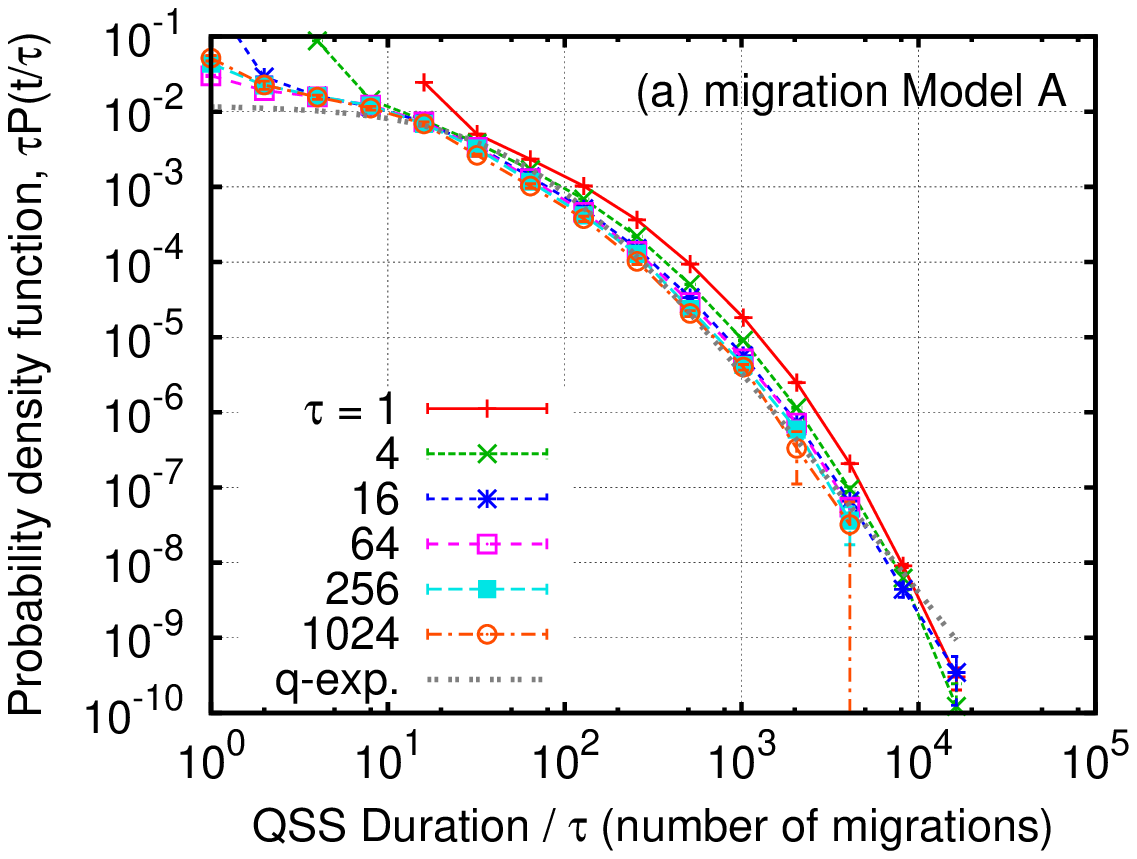}
\label{fig:a-inv-duration-normedtau}
}
\subfigure{
\includegraphics[width=.45\textwidth]{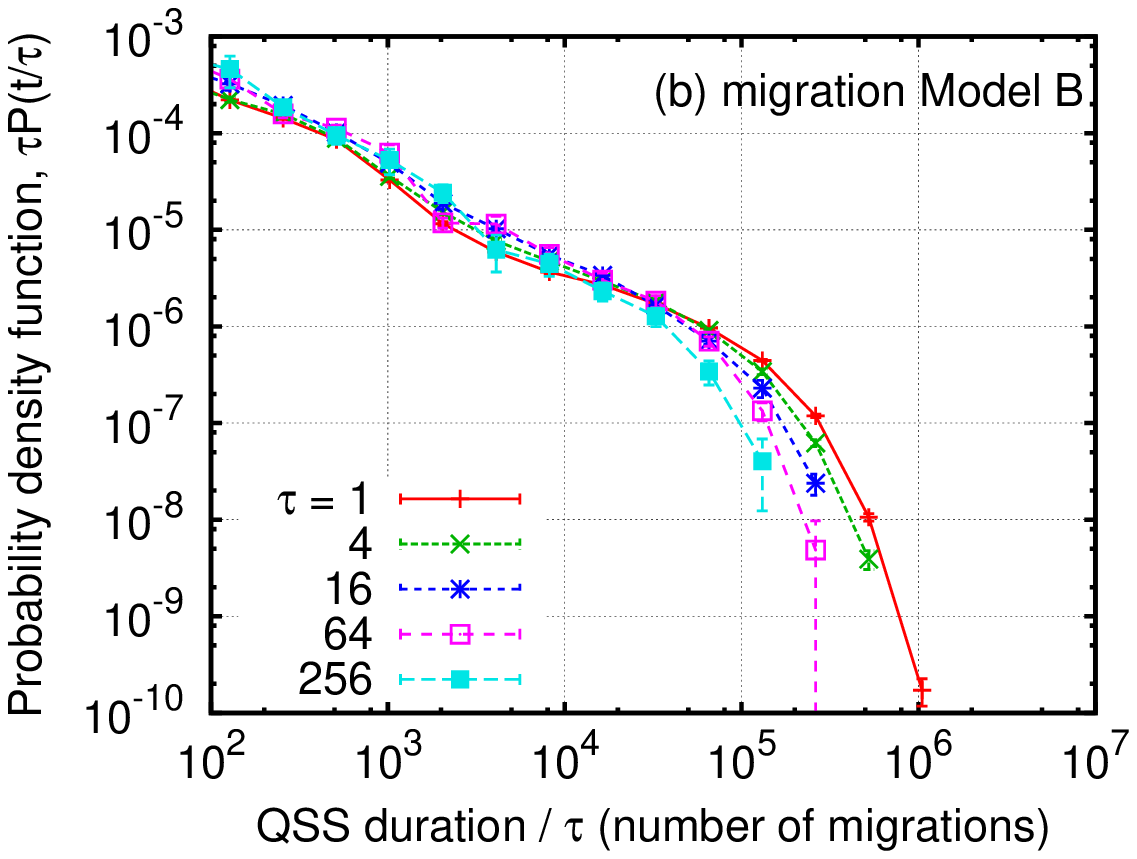}
\label{fig:b-inv-duration-normedtau}
}
\caption{
(Color online) QSS duration distributions normalized by $\tau$, $P(t/\tau)\tau$, 
for (a) migration Model A and (b) migration Model B. 
A $q$-exponential function with $q=1.334$ and $\beta=0.043$ is included in (a) as a guide to the eye.
\label{fig:duration-normedtau}}
\end{center}
\end{figure}

QSS duration distributions normalized by $\tau$ for migration Model A and B are shown in Fig.~\ref{fig:duration-normedtau}. 
The skewed profiles collapse well onto a single curve although the scaling is less clear for Model B. 
We also calculated PSDs for several values of $\tau$ and confirmed that 
the characteristic frequency $f_c$ where the saturation occurs is proportional to $1/\tau$. 
These are consistent with each other. 
Thus, it is confirmed that the time scale for the migration models should be counted by the number of migrations.
This is reasonable since the dynamical systems we studied are asymptotically stable 
and few major extinctions happen without being initiated by the addition of new species. 

\begin{figure}[!ht]
\begin{center}
\includegraphics[width=.45\textwidth]{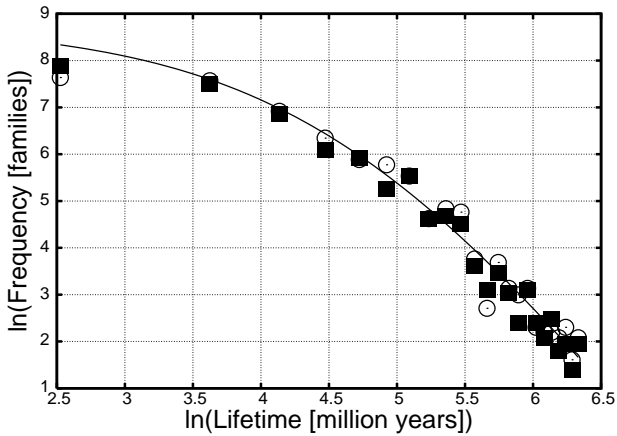}
\caption{
Lifetime distribution of families estimated from fossil records after \citet{Shimada:2003kx}.
The original data consists of the list of presence of each families (present, absent, or unknown) in each geological era.
Open circles and filled boxes represents the estimation by regarding the unknown case as present and absent, respectively.
The line shows a $q$-exponential fit with $q = 1.23$ and $\beta=0.03$.
\label{fig:lifetime-fossil}}
\end{center}
\end{figure}

A possible analytical form for this skewed profile is a $q$-exponential function \citep{Shimada:2003kx}: 
\begin{equation}
	P(t) \propto \left[ 1- (1-q) \beta t \right] ^{1/(1-q)}, 
	\label{eq:q-exponential}
\end{equation}
where $q$ and $\beta$ are fitting parameters. 
The asymptotic form of this $q$-exponential function for $t \rightarrow \infty$ is a simple power-law decay, $t^{1/(1-q)}$.
In the limit of $q \rightarrow 1$, the $q$-exponential function becomes a simple exponential function, $e^{ -\beta t}$. 
The parameter $\beta$ represents the inverse of the characteristic time scale. 
This function is derived from an age-dependent species-mortality function as shown below.
We fitted a $q$-exponential function to the species-lifetime distribution for migration Model A with $\tau=256$ for $t \ge 2\tau$, 
and obtained $q = 1.296(26)$ and $\beta = 0.021(4)$ $[\text{number of migrations}^{-1}]$. 
The agreement of the fitting functions with the simulation data is quite good. 
A previous study \citep{Shimada:2003kx} 
revealed that the lifespan of families in fossil data forms a similar 
``skewed profile,'' and the distribution is in reasonable agreement 
with a $q$-exponential function with $q=1.234(33)$ and 
$\beta = 0.0301(44)$ $[\text{million years}^{-1}]$.
(See Fig.~\ref{fig:lifetime-fossil}.)
We note that $q$ for this simulation model is in reasonable agreement with the $q$ for the fossil data. 
We can compare the species-lifetime distribution obtained from the simulation 
with the fossil data although it may be debatable whether the comparison is valid 
because the fossil data show the lifespan not of species but of families. 
Comparison of $\beta$ gives $\tau = 0.7$ million years for migration Model A, 
and $\tau = 6$ thousand years for migration Model B.
From a fitting to the QSS duration distribution for migration Model A with $\tau=256$ on $t\ge 8\tau$, 
$q=1.334(21)$ and $\beta = 0.043(9)$ $[\text{number of migrations}^{-1}]$ are obtained.
This $q$ is also in reasonable agreement with the $q$ for the fossil data. 
(The distribution for migration Model B is not clear enough for such fitting.)
This pattern is so robust that it is also expected to be observed in other models, 
as indeed it is \citep{PhysRevLett.90.068101,shimada-arob2002,Laird:2006it}. 


The origin of the initial power laws in the species-lifetime distributions should be explained by 
the dynamics of unsuccessful species which fail to have a stable positive population. 
We expect that the population of unsuccessful species is described by a 
Galton-Watson like stochastic branching process 
since the population undergoes a reproductive process. 
The $t^{-2}$ decay is consistent with such a process \citep{pigolotti2005sld}. 

It is known that the $q$-exponential function is obtained as a result of age-dependent mortality \citep{Shimada:2003kx}. 
By integrating Eq.~(\ref{eq:q-exponential}), 
one can obtain the ratio of species that live longer than $t$, $S(t)$, 
which is another $q$-exponential function: 
\begin{equation}
	S(t) = \int_{t}^{\infty} P(t^{\prime}) dt^{\prime} = [ 1- (1-q^{\prime}) \beta^{\prime} t ]^{1/(1-q^{\prime})},
\label{eq:survivorship}
\end{equation}
where $q^{\prime} = 1/(2-q)$ and $\beta^{\prime} = (2-q)\beta$. 
The integral in Eq.~(\ref{eq:survivorship}) converges only if $1 \leq q < 2$, 
and this condition is satisfied for the simulation data. 
The probability that a species which has lived for $t$ generations goes extinct in the $(t+1)$th generation is 
\begin{equation}
	m(t) = \frac{P(t)}{S(t)} = \frac{(2-q)\beta}{1-(1-q)\beta t}.
\end{equation}
Thus the $q$-exponential function is obtained from a mortality function $m(t)$ 
which is the inverse of a linear function of $t$. 
For $\beta t \ll 1$, $m(t) \approx (2-q)\beta$, which is a small constant. 
For $\beta t \gg 1$, $m(t) \approx (2-q)/(1-q)t$, which is inversely proportional to $t$. 
This is reasonable in an ecological sense since 
species that have already existed a long time may be expected to also exist far into the future.

\section{Time series for lower mutation rates}\label{sec:mu-dep}
\begin{figure}[!ht]
\begin{center}
\subfigure{
\includegraphics[width=.45\textwidth]{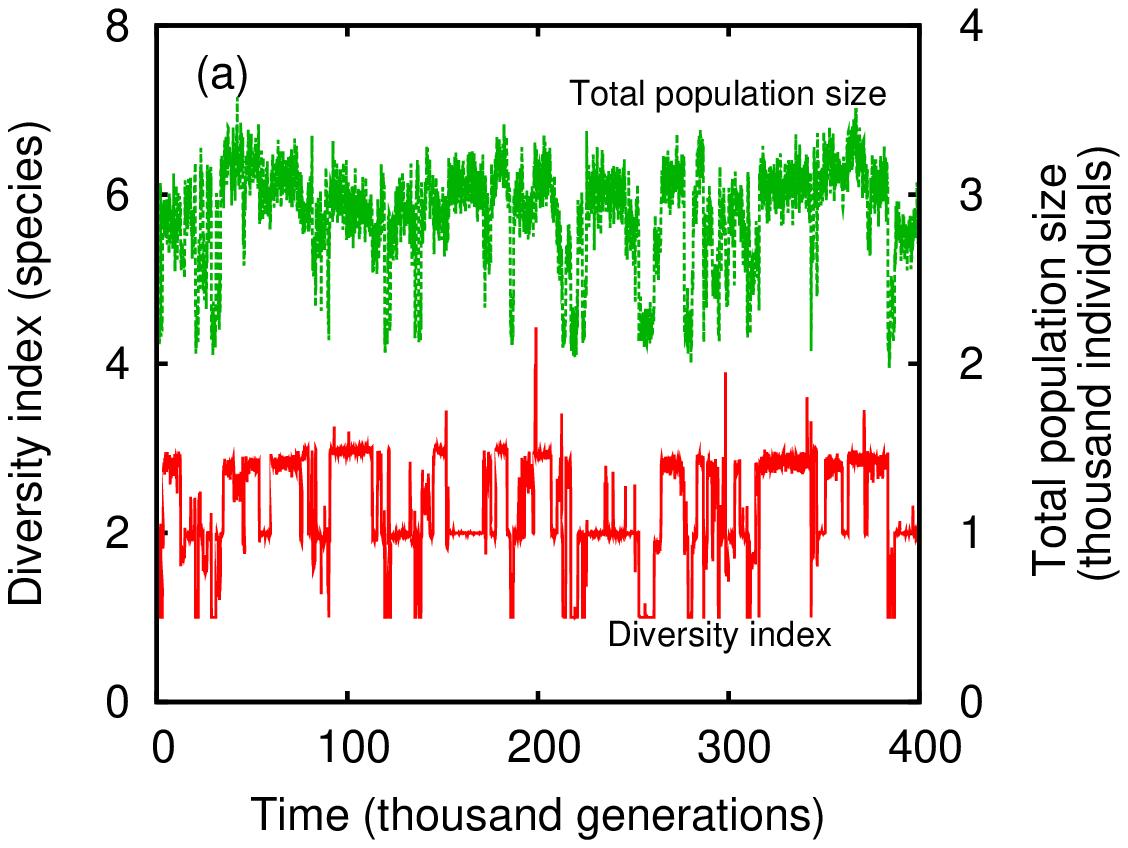}
\label{fig:a-lower-mu}
}
\subfigure{
\includegraphics[width=.45\textwidth]{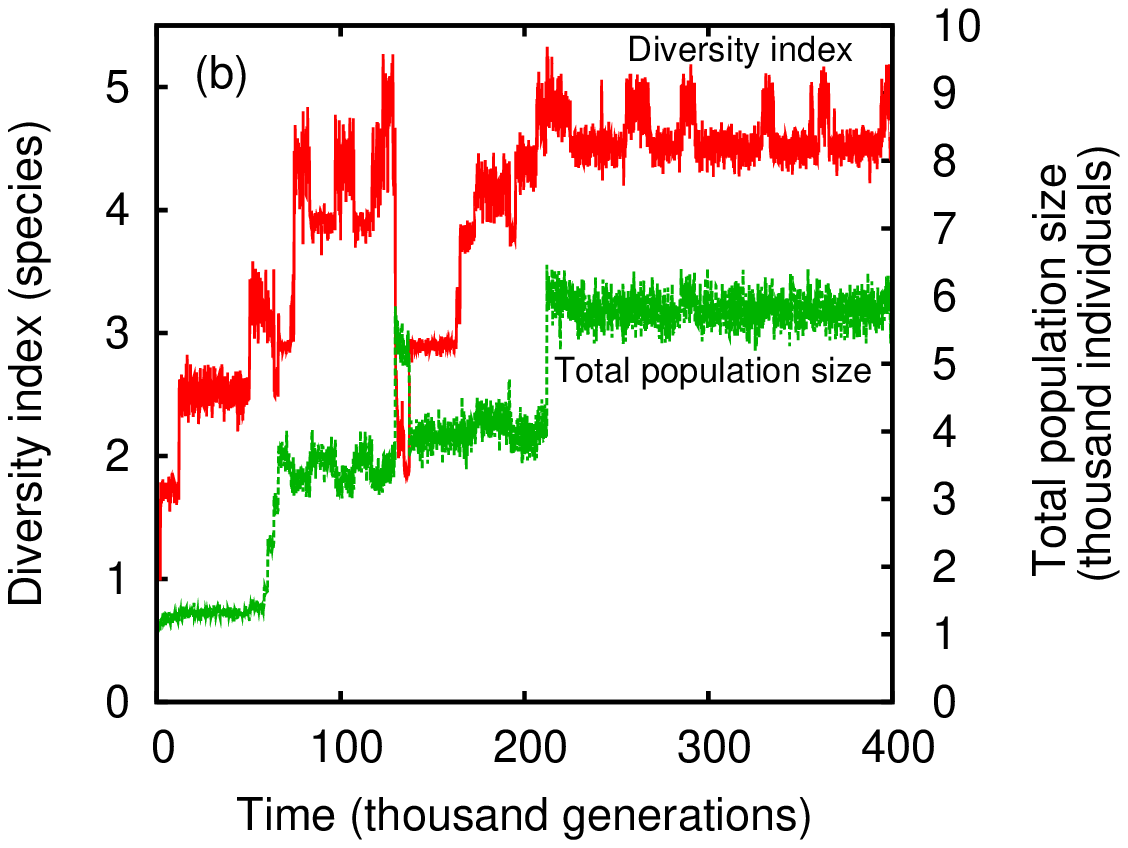}
\label{fig:b-lower-mu}
}
\caption{
  (Color online)
  Typical time series of exponential Shannon-Wiener diversity index and total population size
  for (a) mutation Model A with $\mu=10^{-5}$, $N_0=2000$, and $L=25$ 
  and (b) mutation Model B with $\mu=10^{-5}$, $R=2000$, and $L=24$.
  \label{fig:lower-mu}}
\end{center}
\end{figure}

Although the mutation rates used in the body of the manuscript are high,
the time series are intermittent even with lower mutation rates.
Some examples of the time series with other parameter sets are shown in Fig.~\ref{fig:lower-mu}.
These time series are clearly different from the ones for the migration models.

\end{document}